\documentclass[10pt,twocolumn,twoside]{IEEEtran}
\usepackage{amsmath}
\usepackage{amssymb}
\usepackage{algorithm}
\usepackage{algorithmicx}
\usepackage{algpseudocode}
\usepackage{graphicx}
\usepackage{subfigure}
\usepackage{epstopdf}
\usepackage{cite}
\usepackage{textcomp}
\usepackage{mathrsfs}
\usepackage{color}
\usepackage{booktabs}
\usepackage{cases}
\usepackage{setspace}
\usepackage{bm}
\usepackage{cuted}
\usepackage{booktabs}
\usepackage{stfloats}
\usepackage{makecell}
\usepackage{mathtools}

\makeatletter

\renewcommand{\maketag@@@}[1]{\hbox{\m@th\normalsize\normalfont#1}}%

\makeatother

\begin{document}
	
	\title{\huge Computation-Efficient Backscatter-Blessed MEC with User Reciprocity} 
	
	\author{ %Authors
				Bowen~Gu, Hao~Xie, Dong~Li, \emph{Senior Member, IEEE}
				\IEEEcompsocitemizethanks{			
%					\IEEEcompsocthanksitem This work was supported in part by The Science and Technology Development Fund, Macau SAR, under Grants 0018/2019/AMJ, 0110/2020/A3, and 0029/2021/AGJ.  Bowen Gu and Hao Xie contributed to this work equllly and should be regarded as co-first authors. (\textit{Corresponding author: Dong Li}.)
				\IEEEcompsocthanksitem The authors are with the School of Computer Science and Engineering, Macau University of Science and Technology, Avenida Wai Long, Taipa, Macau 999078, China (e-mails: 21098538ii30001@student.must.edu.mo, 3220005631@student.must.edu.mo, dli@must.edu.mo).} 	\vspace{-10mm}
	}

	%	\markboth{Journal of \LaTeX\ Class Files,~Vol.~14, No.~8, August~2015}%
	%	{Shell \MakeLowercase{\textit{et al.}}: Bare Advanced Demo of IEEEtran.cls for IEEE Computer Society Journals}

\setlength{\textfloatsep}{3pt}

	\maketitle
	\thispagestyle{empty}
	\pagestyle{empty}
	
	\begin{abstract}
		This letter proposes a new user cooperative offloading protocol called user reciprocity in backscatter communication (BackCom)-aided mobile edge computing systems with efficient computation, whose quintessence is that each user can switch alternately between the active or the BackCom mode in different slots, and one user works in the active mode and the other user works in the BackCom mode in each time slot. In particular, the user in the BackCom mode can always use the signal transmitted by the user in the active mode for more data transmission in a spectrum-sharing manner. To evaluate the proposed protocol, a computation efficiency (CE) maximization-based optimization problem is formulated by jointly power control, time scheduling, reflection coefficient adjustment, and computing frequency allocation, while satisfying various physical constraints on the maximum energy budget, the computing frequency threshold, the minimum computed bits, and harvested energy threshold. To solve this non-convex problem, Dinkelbach's method and quadratic transform are first employed to transform the complex fractional forms into linear ones. Then, an iterative algorithm is designed by decomposing the resulting problem to obtain the suboptimal solution. The closed-form solutions for the transmit power, the RC, and the local computing frequency are provided for more insights. Besides, the analytical performance gain with the reciprocal mode is also derived. Simulation results demonstrate that the proposed scheme outperforms benchmark  schemes regarding the CE. 
	\end{abstract}
	\IEEEpeerreviewmaketitle
	\section{Introduction}
	
    Over recent years,  the Internet-of-Things (IoT) and mobile communication technologies have made significant stride in popularizing and implementing, creating  varieties of products and applications that take into account what users need and want \cite{LiuLiDai}.  However, some challenges are coming on the heels. The irreconcilable contradiction between users' ever-increasing data needs and their insufficient computing abilities and battery capacities is a case in point.
    
    In this regard, mobile edge computing (also called multi-access edge computing) (MEC) is a viable paradigm to grapple with this issue, which enables IoT devices to execute their computation-intensive and latency-sensitive applications at the vicinal edge servers \cite{DjigalXuLiu}. 
     %However, how to efficiently execute these applications of users via MEC is a crucial issue to be faced with, since not only massive tasks need to be properly offloaded and scheduled to edge servers, but also the quality-of-experience/service (QoE/S) requirements of users should be guaranteed. To contend with this, one of the effective solutions is to investigate the suitable resource allocation scheme for the computing and offloading policy for different users in different networks, which has already attracted much attention. Specifically, for offloading policy, binary- and partial-offloading modes are successively proposed to better complete the task executions for users \cite{ZhouHu}. Adopting which of them depends on not only the data size and the required latency of each task but also its data type (i.e., whether task data can be divided into different groups or not). Besides, there have been fruitful works devoted to the transmission strategies for performance improvement  in various scenarios including, e.g.,  wireless-powered communication networks \cite{ZhouHu}, reconfigurable intelligent surface (RIS)-aided systems \cite{XieGuLi}, vehicular networks \cite{ZhangLiu}, etc.
	However, although computation offloading provides stress relief for users, it also costs quite high energy consumption for data transmission, which is still overwhelming for energy-hungry users. A turnaround in this deadlock comes with the advent of the backscatter communication (BackCom), which relies on no recourse to the active radio-frequency components to transmit information, but reflects the incident signal from the power source (see, e.g., \cite{Gu2022,Lixingwang}). Recognizing its superior performance regarding low power consumption, some contributions have been made to MEC systems by leveraging BackComs. For instance, the traditional active transmission was replaced by BackCom in \cite{XuGuHuLi}  for green MEC. 
	The bidirectional Bluetooth BackCom was explored in \cite{JiangGong} for fast and reliable transmission to the edge server.
	To further enhance the offloading efficiency, the hybrid backscatter/harvest-then-transmit protocol was adopted in  \cite{YeShiChuHu,HeWuHe, ZargariTrllambura}, where the multi-user active/passive switching \cite{YeShiChuHu}, the user cooperation in a relaying manner \cite{HeWuHe}, and the RIS-assisted offloading strategy \cite{ZargariTrllambura} were studied, respectively.
    
    Albeit, the works mentioned are not out of the inherent flaws of BackComs, which rely heavily on (dedicated/ambient) power sources (PSs) to supply energy for data transmission. Since, in practice, the PSs are not always available. To surmount this obstacle, we take a cue from \cite{KangLiangYang,Li3}, in which the passive users are permitted to coexist with existing active users/systems in a spectrum-sharing manner. With this in mind, in this letter, we investigate the switchable active-passive offloading protocol for two users by leveraging a user-cooperation manner, in which when one user is working in the active mode, another user is able to work in the BackCom mode by using the radiation signal of the active user. Then, by switching their roles in different slots, these two users can help each other to complete their data transmission, which we call \textit{user reciprocity}. Notably, although user cooperation is also covered in \cite{HeWuHe}, which most closely matches this work, it considers one user (near to PS) acts as a helper to relay the information of another user (far to PS). However, one user always serves as the helper, which leads to potential unfairness between two users. Besides, the fixed cooperation mode between users may not contribute most to the CE performance. This motivates us to investigate a new user cooperation protocol, which is utterly different from \cite{HeWuHe}.
   It should also be noted that the considered active-passive user reciprocity can also be regarded as the hybrid active and passive mode but for two separate users, which is different from \cite{Li1,Li2} for the integrated functionality for one user. The main contributions are summarized as follows
	\begin{itemize}
		\item To evaluate the performance of the proposed protocol, an optimization problem with the computation efficiency (CE) maximization is formulated by jointly optimizing the transmit power, the offloading time, the reflection coefficient (RC), and the local computing frequency, where the maximum energy budget, the computing frequency threshold, the minimum computed bits, and harvested energy threshold are considered.

      \item To solve this knotty problem, Dinkelbach's method and quadratic transform are employed to transform the intricate fractional forms into linear ones. Then,  an alternating algorithm is designed after decomposing the resulting problem. For analytical insights, we not only provide the closed-form solutions for the transmit power, the RC, and the local computing frequency, but also derive the analytical performance gain with reciprocal mode.

      \item Simulation results demonstrate that the proposed scheme is better than benchmark schemes in terms of CE by evaluating the impact of the maximum energy budget and the minimum computed bits on the system performance. 
	\end{itemize}

	\section{System Model}
	
	\begin{figure}[t]
%		\vspace{-5mm}
		\centerline{\includegraphics[width=3.5in]{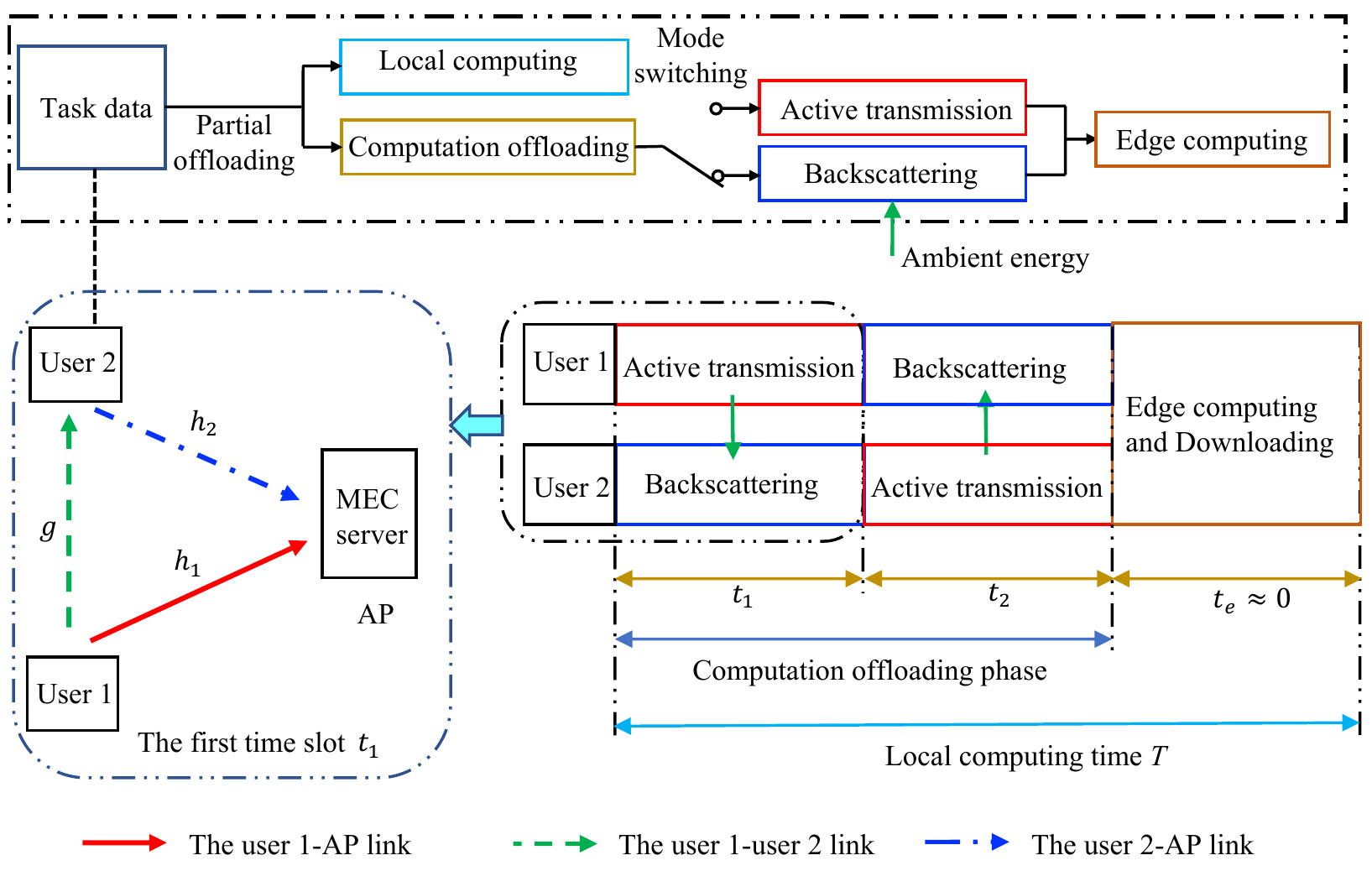}}
		\caption{An MEC-based network with BackCom assistance.}
		\label{fig1}
	\end{figure} 

\subsection{System structure and transmission mechanism}
An MEC-based system with BackCom assistance is considered, as shown in Figure 1, which consists of one access point (AP)  and two users. Specifically, the AP is equipped with an MEC server that can remotely execute users' tasks, and each user can switch between active and backscattering modes. Assume that each task is bit-independent and thus can be processed simultaneously by edge computing and local computing with partial offloading. Moreover,  the quasi-static flat-fading channel is further assumed in this system, where the channel state information remains constant but may change from one time slot to another.

The frame structure consists of two stages, the computation offloading phase and the edge computing and downloading phase. The frame duration is denoted by $T$, which is less than the latency requirements of computation tasks. Besides, we ignore the time spent on edge computing and downloading due to the strong computing capability of the MEC server and the small size of computation results (i.e., $t_e\approx0$). During the computation offloading phase, users takes turns to actively transmit their data to the AP, obeying a time-division-multiple-access manner. Meanwhile, another user can utilize the signal radiated by the current active user to perform BackCom. 
 
 \subsection{Energy supply for backscattering}
 When the $k$-th user ($k\in \{1,2\}$) is actively transmitting data for offloading, the received signal at the $j$-th user ($j=3-k$) can be expressed as 
% \begin{equation} \label{b1}
 	$r_j=\sqrt{p_k}gs_k+n_{\text{u}}$,
% \end{equation}
 where $p_k$ denotes the transmit power of the $k$-th user, $g$ is the channel coefficient of interuser link, $s_k$ is the transmit symbol of the $k$-th user, satisfying $\mathbb{E}(|s_k|^2)=1$, and $n_u$ is the noise power. Note that the noise is  weak and can be ignored for EH and backscattering transmission. Therefore,  the harvested energy of the $j$-th user can be expressed as 
  \begin{equation} \label{b2} \small 
 	E_j^{\text{h}}=t_k \zeta (1-\alpha_j) p_k|g|^2,
% 	E_j^{\text{h}}=t_k \min (\zeta (1-\alpha_j) p_k|g|^2,P_k^{\text{sat}}),
 \end{equation}
where $\zeta$ denotes the EH coefficient, $\alpha_j$ is the RC of the $j$-th user, $t_k$ is the $k$-th time slot. 
%, and $P_k^{\text{sat}}$ is the saturation power for EH.

Thus, to perform backscattering, the following condition has to be satisfied, i.e., 
%\begin{equation}  \label{b3} 	
		$E_j^{\text{h}}\ge t_kP_j^{\text{bc}}$,
%\end{equation}
where $P_j^{\text{bc}}$ denotes the circuit power for backscattering of the $j$-th user.

\subsection{Partial Offloading}
  \subsubsection{Computation offloading} 
According to the above illustrations, in each time slot, there are always one active user and one passive user jointly offloading data. Thus, in the $k$-th slot, the received hybrid signal at the AP can be expressed as 
% \begin{equation} \label{d1}
 	$y_k=\sqrt{p_k}h_ks_k+\sqrt{\alpha_jp_k}h_jgs_ks_j+n_{\text{a}}$,
% \end{equation}
where $h_k$ is channel coefficient of the user $k$-AP link,   and $n$ is the background noise, satisfied $n_{\text{a}}\sim \mathcal{CN}(0,\sigma^2)$. 

Recalling $y_k$, $s_k$ and $s_j$ interfere with each other. To improve the transmission performance, successive interference cancellation is applied for detecting $s_k$ and $s_j$. Since the backscattering power $s_j$ is much lower than that of  $s_k$, we first detect
 $s_k$ from $y_k$. Thus, the received signal-tointerference-plus-noise-ratio (SINR) of $s_k$ can be expressed as
  \begin{equation} \label{d2} \small 
 	\gamma_k^{\text{a}}=\frac{p_k|h_k|^2}{\alpha_jp_k|h_j|^2|g|^2+\sigma^2}.
 \end{equation}

Therefore, the active offloading data size of the $k$-th user can be expressed as 
%  \begin{equation} \label{d3}
$R_k^{\text{a}}=t_kB\log_2(1+\gamma_k^{\text{a}}),$
%\end{equation}
where $B$ is the communication bandwidth.

On the other hand, when $s_k$ is detected at the AP, it can be removed from $y_k$, and thus the interference caused by  $s_k$ can be completely canceled. Then, the received signal-to-noise-ratio (SNR) of $s_j$ can be expressed as
  \begin{equation} \label{d4} \small 
	\gamma_j^{\text{b}}=\frac{\alpha_jp_k|h_j|^2|g|^2}{\sigma^2}, {\text{if}}~\gamma_k^{\text{a}}\ge \gamma_k^{\text{th}}, 
\end{equation}
where $\gamma_k^{\text{th}}$ is the SINR threshold for detecting $s_k$ from $y_k$.

Accordingly, the backscattering offloading data size of the $j$-th user can be expressed as 
%\begin{equation} \label{d5}
	$R_j^{\text{b}}=t_kB\log_2(1+\gamma_j^{\text{b}}).$
%\end{equation}

\subsubsection{Local computing} 

It is noted that each user can perform local computing in the entire frame. Defining $C_k$ as the CPU cycles required for computing one bit of data, the number of locally computed bits at the $k$-th user can be expressed as
\begin{equation} \label{l1} \small 
	R_k^{\text{l}}=\frac{Tf_k}{C_k},
\end{equation}
where $f_k$ denotes the CPU frequency of the $k$-th user.

The CPU architecture of each user adopts the advanced dynamic frequency and voltage scaling technique, the energy consumption of the $k$-th user used for local computing is
\begin{equation} \label{l2} \small 
	E_k^{\text{l}}=T\phi_k f_k^3,
\end{equation}
where $\phi_k$ is the switched capacitance coefficient \cite{XuGuHuLi}.

\subsection{Computation efficiency}
\subsubsection{Computation bits}For the $k$-th user, the total number of computed bits mainly includes two parts, i.e., the local and the offloaded parts, which can be expressed as 
\begin{equation} \label{c1} \small 
	R_k^{\text{tot}}=\frac{Tf_k}{C_k}+R_k^{\text{a}}+R_k^{\text{b}},
\end{equation}
where $R_k^{\text{b}}$ is the backscattering transmission bits at the $j$-slot, which can be expressed as $R_k^{\text{b}}=t_jB\log_2(1+\frac{\alpha_kp_j|h_k|^2|g|^2}{\sigma^2})$.
\subsubsection{Energy consumption} Since the energy used for BackCom of the $k$-th user is from the $j$-th user, and the harvested energy is enough to cover the circuit consumption of BackCom. Therefore, for the $k$-th user, the consumed energy mainly comes from the active transmission and local computing, which can be expressed as 
\begin{equation} \label{c2} \small 
	E_k^{\text{tot}}=t_k(p_k+P_k^{\text{ac}})+E_k^{\text{l}},
\end{equation}
where $P_k^{\text{ac}}$ is the constant circuit power consumption for active transmission of the $k$-th user.

Based on the above analysis, the total CE of users, which is defined as the total number of computation bits related to users per Joule, can be expressed as 
\begin{equation} \label{c3} \small 
	\eta=\frac{\sum_{k=1}^2 R_k^{\text{tot}}}{\sum_{k=1}^2 E_k^{\text{tot}}}.
\end{equation}

	\section{Problem Formulation and Algorithm Design}
\subsection{Problem formulation}
Our purpose to maximize the total CE for the reciprocal users by jointly optimizing the transmit power of users, the computation offloading time, the RC for backscattering, and the CPU frequency of users for local computing. Accordingly, the optimization problem is mathematically formulated as
		\begin{equation}  \label{p1}  \small 
	\begin{aligned}	
		& \underset{p_k,  t_k, \alpha_k, f_k}{\mathop{\max }}\, \eta \\
		&\text{s.t.}~
		C_1: 0<\alpha_k\le 1, \forall k,~C_2: \sum_{k=1}^2 t_k\le T, \\
		&\quad ~~C_3:  f_k\le f_k^{\max}, \forall k,~C_4:E_k^{\text{h}}\ge t_jP_k^{\text{bc}}, \forall k, \\
		&\quad ~~C_5: \gamma_k^{\text{a}}\ge \gamma_k^{\text{th}}, \forall k,~C_6: R_k^{\text{tot}}\ge R_k^{\text{th}}, \forall k,  \\
		&\quad ~~C_7: E_k^{\text{tot}} \le E_k^{\text{bud}}, \forall k, \\
	\end{aligned}
\end{equation}  
where $f_k^{\max}$, $R_k^{\text{th}}$, and $ E_k^{\text{bud}}$ are the the maximum CPU frequency,  the minimum required number of computed bits, and maximum energy budget of
the $k$-th user, respectively. In problem (\ref{p1}), $C_1$, $C_2$, and $C_3$ denote the constraints of the RC, the offloading time, the CPU frequency, respectively. 
$C_4$ limits the consumed energy used for BackCom to be less than the harvested energy. $C_5$ ensures that the resulting SINR should be greater than its required threshold. $C_6$ guarantees that the number of computed bits is not less than the required threshold. Finally, $C_7$ limits that the energy used by the user for task execution does not exceed its budget.

\subsection{Problem transformation}
It can be seen that, there are many obstacles in solving  problem (\ref{p1}), one of which is that the non-smooth objective function. To deal with that, Dinkelbach's method is employed. Accordingly, problem (\ref{p1}) can be transformed into 
		\begin{equation}  \label{p2} \small 
	\begin{aligned}	
		& \underset{p_k,  t_k, \alpha_k, f_k}{\mathop{\max }}\, {\sum_{k=1}^2 R_k^{\text{tot}}}-\eta {\sum_{k=1}^2 E_k^{\text{tot}}} \\
		&\text{s.t.}~
		C_1\sim C_7.
	\end{aligned}
\end{equation}  
Yet despite that, the fractional optimization structure regarded to SINR of $R_k^{\text{a}}$ still hinders the solution of problem (\ref{p2}). To break this blockade, the quadratic transform is adopted \cite{ShenYu}. Accordingly,  $R_k^{\text{a}}$  can be rewritten as
  \begin{equation} \label{t1} \small 
	\bar R_k^{\text{a}}{=}t_kB\log_2(1+2y_k\sqrt{p_k|h_k|^2}-y_k^2({\alpha_jp_k|h_j|^2|g|^2+\sigma^2})),
\end{equation}
where $y_k$ is auxiliary variable introduced by the quadratic transform for the $k$-th user.

According to (\ref{t1}), problem (\ref{p2}) can be transformed into
		\begin{equation}  \label{p3}  \small 
	\begin{aligned}	
		& \underset{p_k,  t_k, \alpha_k, f_k, y_k}{\mathop{\max }}\, \sum_{k=1}^2 \left( {\frac{Tf_k}{C_k}}+{\bar R_k^{\text{a}}+R_k^{\text{b}}}\right) -\eta {\sum_{k=1}^2 E_k^{\text{tot}}} \\
		&\text{s.t.}~
		C_1\sim C_5, C_7, C_{6a}: {\frac{Tf_k}{C_k}}+{\bar R_k^{\text{a}}+R_k^{\text{b}}}\ge R_k^{\text{th}}, \forall k. \\
	\end{aligned}
\end{equation} 

\subsection{Algorithm design}
So far, problem (\ref{p3}) has become more tractable than problem (\ref{p2}), but it is still non-convex and hard to obtain the optimal solution due to the strong coupled relationship among variables. Thanks to the alternating optimization method, we decompose this problem into two subproblems as follows.
\subsubsection{Subproblem A and solution} With the fixed $t_k$ and $y_k$, defining $\bar p_{k,j}=\alpha_kp_j$ and $\hat p_{j,k}=\alpha_jp_k$,  and introducing a slack variable $\bar \gamma_k^{\text{a}}$, problem (\ref{p3}) can be rewritten as 
		\begin{equation}  \label{p3a}  \small 
	\begin{aligned}	
		& \underset{p_k, \bar p_{k,j}, \hat p_{j,k}, f_k, \bar \gamma_k^{\text{a}}}{\mathop{\max }}\, \sum_{k=1}^2 \left( {\frac{Tf_k}{C_k}}+{\hat R_k^{\text{a}}+\bar R_k^{\text{b}}}\right) -\eta {\sum_{k=1}^2 E_k^{\text{tot}}} \\
		&\text{s.t.}~
		C_3, C_7, C_{1a}: 0\le \bar p_{k,j}\le p_j,  0\le \hat p_{k,j}\le p_k, \forall k, \\
		&\quad ~~C_{4a}: t_j\zeta (p_j-\bar p_{k,j}) |g|^2\ge t_jP_k^{\text{bc}}, \forall k,\\
		&\quad ~~C_{5a}: {p_k|h_k|^2}\ge (\hat p_{j,k}|h_j|^2|g|^2+\sigma^2)\gamma_k^{\text{th}}, \forall k,\\
		&\quad ~~C_{6b}: {\frac{Tf_k}{C_k}}+{\hat R_k^{\text{a}}+\bar R_k^{\text{b}}}\ge R_k^{\text{th}}, \forall k, \\
		&\quad ~~C_{8}: 2y_k\sqrt{p_k|h_k|^2}-y_k^2({\hat p_{j,k}|h_j|^2|g|^2+\sigma^2})\ge \bar \gamma_k^{\text{a}},\forall k, 
	\end{aligned}
\end{equation} 
where $\hat R_k^{\text{a}}=t_kB\log_2(1+\bar \gamma_k^{\text{a}})$ and $\bar R_k^{\text{b}}=t_jB\log_2(1+\frac{\bar p_{k,j}|h_k|^2|g|^2}{\sigma^2})$.

It can be observed that problem (\ref{p3a}) is a convex optimization problem, which can be solved by many standard convex methods, such as CVX. Here, in order to obtain more insights, the Lagrange dual method is applied to derive the closed-form solution for some optimization variables in problem (\ref{p3a}). Specifically, the Lagrange function of problem (\ref{p3a}) is formulated as (\ref{s1}), where ${\bf{\Xi}}_k=\{p_k, \bar p_{k,j}, f_k, \bar \gamma_k^{\text{a}}, \psi_k, \mu_k, \varepsilon_k, \lambda_k, \omega_k, \chi_k, \varkappa_k\}$, $\psi_k, \mu_k, \varepsilon_k, \lambda_k, \omega_k$, $\chi_k$, and $\varkappa_k$ are non-negative Lagrange multipliers, which can be  updated via the gradient based method.

\begin{table*}[t]
	\vspace{-5mm}
	\small 
	\setcounter{equation}{13}
	\begin{equation} \label{s1} \small 
		\begin{aligned}	
			\mathcal{L}{({\bf{\Xi}}_k)}{=}&\sum_{k=1}^2 \left( {\frac{Tf_k}{C_k}}{+}{\hat R_k^{\text{a}}{+}\bar R_k^{\text{b}}}\right){-}\eta {\sum_{k=1}^2 E_k^{\text{tot}}}{+}\sum\limits_{k=1}^2 \psi_k  (p_j{-}\bar p_{k,j}){+}\sum\limits_{k=1}^2\mu_k(f_k^{\max}{-} f_k)
			{+}\sum_{k=1}^2\varepsilon_k(t_j\zeta (p_j{-}\bar p_{k,j}) |g|^2){-} t_jP_k^{\text{bc}})\\
			&{+}\sum\limits_{k=1}^2 \chi_k( E_k^{\text{bud}}- E_k^{\text{tot}}){+}\sum\limits_{k=1}^2 \lambda_k({p_k|h_k|^2}{-} (\hat p_{j,k}|h_j|^2|g|^2{+}\sigma^2)\gamma_k^{\text{th}})  {+}\sum\limits_{k=1}^2 \omega_k \left( {\frac{Tf_k}{C_k}}{+}{\hat R_k^{\text{a}}{+}\bar R_k^{\text{b}}}{-}R_k^{\text{th}}\right)\\	
			&{+}\sum\limits_{k=1}^2 \varkappa_k(2y_k\sqrt{p_k|h_k|^2}-y_k^2({\hat p_{j,k}|h_j|^2|g|^2+\sigma^2})- \bar \gamma_k^{\text{a}}).
		\end{aligned}
	\end{equation}  
	\hrule 
%		\vspace{1mm}
%			\begin{equation}\label{s2} 
%		\begin{aligned}
%			%						&\frac{\partial \mathcal{L}({\bf{\Xi}}_k)}{\partial p_k}{=}\lambda_k(|h_k|^2-\alpha_j|h_j|^2|g|^2\gamma_k^{\text{th}})+\varkappa_k(\frac{y_k|h_k|^2}{\sqrt{p_k|h_k|^2}}-y_k^2\alpha_j|h_j|^2|g|^2)-(\eta+\chi_k)t_k=0\\
%			%						&\frac{y_k|h_k|^2}{\sqrt{p_k|h_k|^2}}\varkappa_k=(\eta+\chi_k)t_k-\lambda_k(|h_k|^2-\alpha_j|h_j|^2|g|^2\gamma_k^{\text{th}})+y_k^2\alpha_j|h_j|^2|g|^2\varkappa_k\\
%			%						&{\sqrt{p_k|h_k|^2}}=\frac{y_k|h_k|^2\varkappa_k}{(\eta+\chi_k)t_k-\lambda_k(|h_k|^2-\alpha_j|h_j|^2|g|^2\gamma_k^{\text{th}})+y_k^2\alpha_j|h_j|^2|g|^2\varkappa_k}\\
%%			&\left( \frac{y_k|h_k|\varkappa_k}{(\eta{+}\chi_k)t_k{-}\lambda_k(|h_k|^2{-}\alpha_j|h_j|^2|g|^2\gamma_k^{\text{th}}){+}y_k^2\alpha_j|h_j|^2|g|^2\varkappa_k}\right) ^2.\\
%			&p_k^*{=}\left( \frac{y_k|h_k|\varkappa_k}{(\eta{+}\chi_k)t_k{-}\lambda_k|h_k|^2+\alpha_j|h_j|^2|g|^2(\lambda_k\gamma_k^{\text{th}}+y_k^2\varkappa_k)}\right) ^2.
%		\end{aligned}
%	\end{equation}
%	\hrule 
	\vspace{1mm}
	\setcounter{equation}{20}
	\begin{equation}\label{rm1} \small 
		C_{k}^{\text{tot}}{=}\frac{T}{2}B\left( \log_2\left( 1+\frac{P_0|h_k|^2}{\alpha_jP_0|h_j|^2|g|^2+\sigma^2}\right)+\log_2\left( 1+\frac{\alpha_jP_0|h_j|^2|g|^2}{\sigma^2}\right)\right){=}\frac{T}{2}B\log_2\left( 1+\frac{\alpha_jP_0|h_j|^2|g|^2+P_0|h_k|^2}{\sigma^2}\right). 
	\end{equation}
	\hrule 
	\vspace{1mm}
	\setcounter{equation}{23} 	
	\begin{equation}\label{rm3} \small 	
		C_{k}^{\text{gap}}=C_{k}^{\text{tot}}-\bar C_{k}^{\text{tot}}= \left\{ \begin{aligned}
			&\frac{T}{2}B\log_2\left( 1+\frac{(P_0|g|^2-p_j^{\text{bc}}/\zeta)|h_j|^2}{P_0|h_k|^2+\sigma^2}\right), ~~~\text{Case A:} 0\le 1-p_j^{\text{bc}}/\zeta P_0|g|^2\le  \frac{P_0|h_k|^2-\gamma_k^{\text{th}}\sigma^2}{\gamma_k^{\text{th}}P_0|h_j|^2|g|^2} ,\\ 
			&\frac{T}{2}B\log_2\left( 1+\frac{P_0|h_k|^2-\gamma_k^{\text{th}}\sigma^2}{(P_0|h_k|^2+\sigma^2)\gamma_k^{\text{th}}}\right), ~~~~~~~~\text{Case B:} 0\le 1-p_j^{\text{bc}}/\zeta P_0|g|^2>  \frac{P_0|h_k|^2-\gamma_k^{\text{th}}\sigma^2}{\gamma_k^{\text{th}}P_0|h_j|^2|g|^2}.\\ 
		\end{aligned} \right.
	\end{equation}
	\hrule 
\end{table*}

According to KKT conditions, we have the corresponding closed-form solutions, which can be expressed as follows
	\setcounter{equation}{14}
				\begin{equation}\label{s2} 
		\begin{aligned}
p_k^*{=}\left( \frac{y_k|h_k|\varkappa_k}{(\eta +\chi_k)t_k{-}\lambda_k|h_k|^2}\right)^2, \forall k, 
		\end{aligned}
	\end{equation}
\begin{equation}\label{s6}  \small 
{\bar p_{k,j}^*}{=}\left[\frac{t_jB(1+\omega_k)}{\ln 2(\varepsilon_kt_j\zeta |g|^2{+}\psi_k)}{-}\frac{\sigma^2}{|h_k|^2|g|^2}\right]^{+}, \forall k,
\end{equation}
\begin{equation}\label{s7}  \small 
	f_k^*=\sqrt{\frac{(1+\omega_k)\frac{T}{C_k}-\mu_k}{\eta3T\phi_k}}, \forall k, \\
\end{equation}
\begin{equation}\label{s8}  \small 
	(\bar \gamma_k^{\text{a}})^*=\left[\frac{t_kB(1+\omega_k)}{\ln 2\varkappa_k}{-}1\right]^{+}, \forall k,
\end{equation}
where $[x]^+=\max(0,x)$.

\textbf{\textit{Remark 1:}} When the $k$-th user is in the active mode, it can be seen from (\ref{s2}), more $p_k^*$ will be used for computation offloading at the $k$-slot, when $|h_k|^2$ is higher. Since the SINR for active transmission can be enhanced with such a channel gain. From (\ref{s6}), when the $k$-th user is in BackCom mode, a good  $|h_k|^2$ can also make the backscattering power (i.e., $\alpha_kp_j$). It is noted that  $|h_k|^2\ge \frac{\ln 2(\varepsilon_kt_j\zeta |g|^2{+}\psi_k)\sigma^2}{t_jB(1+\omega_k)|g|^2}$ should be satisfied to ensure BackCom can be processed for the $k$-th user. Moreover, when the $k$-th user is computing locally, we can know from (\ref{s7}), $f_k^*$ is inversely proportional to $\eta$, indicating that the user can moderately reduce the bits for local computing to boost the system CE. According to (\ref{s2}) and (\ref{s8}), we find that $\varkappa_k>0$ always holds, which means that the relaxation in $C_8$ is tight when the optimal solution is obtained.

Besides, when $\alpha_k$, $f_k$, and $p_k$ are obtained, optimal $y_k$ can be found in closed
form as
\begin{equation} \label{y1} \small 
	y_k^*=\frac{\sqrt{p_k|h_k|^2}}{\alpha_jp_k|h_j|^2|g|^2+\sigma^2}, \forall k.
\end{equation}

\subsubsection{Subproblem B and solution} With the fixed $\alpha_k$, $f_k$, $p_k$, and $y_k$,  the remaining problem for time allocation can be rewritten as 
		\begin{equation}  \label{p3b}  \small 
	\begin{aligned}	
		& \underset{t_k}{\mathop{\max }}\, \sum_{k=1}^2 \left( {\frac{Tf_k}{C_k}}+{\bar R_k^{\text{a}}+R_k^{\text{b}}}\right) -\eta {\sum_{k=1}^2 E_k^{\text{tot}}} \\
		&\text{s.t.}~
		C_2, C_4, C_{6a}, C_7. \\
	\end{aligned}
\end{equation} 
It is obvious that problem (\ref{p3b}) is a linear problem w.r.t. $t_k$, which can be solved by the linear programming method.

\subsection{Performance improvement with reciprocal mode}
In this subsection, we gain more insight into the proposed user reciprocity operating in the hybrid active and passive mode. Specifically, for ease of analysis, the full-offloading policy is adopted. Besides, we further assume that the average time policy is considered during the computation offloading phase and that each user transmits data in a constant power, i.e., $t_k=T/2$, and $p_k=P_0$. Our purpose is to explore the gap of the computed bits with the same time and power under the two different modes, i.e., reciprocal mode (with BackCom) and non-reciprocal mode (without BackCom).
\subsubsection{Reciprocal mode}The total number of computed bits of the $k$-user can be rewritten as (\ref{rm1}), where $\alpha_j$ is given by 
\setcounter{equation}{21}
\begin{equation}\label{rm1a}  \small 
	\alpha_j=\max\left\lbrace 0,\min \left( 1-p_j^{\text{bc}}/\zeta P_0|g|^2, \frac{P_0|h_k|^2-\gamma_k^{\text{th}}\sigma^2}{\gamma_k^{\text{th}}P_0|h_j|^2|g|^2}\right) \right\rbrace. 
\end{equation}

\subsubsection{Non-reciprocal mode} The total number of computed bits of the $k$-user can be rewritten as 
\begin{equation}\label{rm2} \small 
	\bar C_{k}^{\text{tot}}=\frac{T}{2}B \log_2\left( 1+\frac{P_0|h_k|^2}{\sigma^2}\right).
\end{equation}
\subsubsection{Performance gap} The gap of the total number of computed bits of the $k$-user can be rewritten as (\ref{rm3}).

\textbf{\textit{Remark 2:}} It can be seen from (\ref{rm3}), when Case A is applicable, this allows the conclusion that if $|g|^2>p_j^{\text{bc}}/P_0\zeta$ is satisfied, the transmission performance of the proposed mode is always better than that of the traditional mode. On the other hand, when Case B holds, we can obtain the the same conclusion, if $|h_k|^2>\gamma_k^{\text{th}}\sigma^2/P_0$ is met. Importantly, these results provide compelling evidence for the fact that BackCom will be beneficial regarding the computation offloading. More general comparisons between them can be seen in simulation results.

\section{Simulation Results} 
In this section, we provide simulation results to evaluate the performance of the proposed scheme.  The distances between the AP and users  within 10 meters, and the distance among different BDs is within 4 meters. The distance-dependent pass loss is modeled by $P_L=d^{-\beta}$, in which $d$ denotes the Euclidean distance in meters and $\beta=2.2$ is the path loss exponent. Besides, Rician fading is applied as small-scale fading for all channels, where the Rician factor is 2.8 dB.  Other parameters include $T=1$s, $C_k=1000$cycles/bit, $\phi_k=10^{-26}$, $f_k^{\max}=1$GHz, $E_k^{\text{bud}}=1$J, $P_k^{\text{bc}}=0.1$mW, $P_k^{\text{ac}}=10$mW, $\zeta=0.8$, $B=0.1$MHz, $\gamma_k^{\text{th}}=100$, $R_k^{\text{th}}=0.2$Mbits, and $\sigma^2=10^{-8}$ mW. For scheme comparison, we define some benchmark schemes, such as full-offloading scheme, full-local computing scheme, and non-reciprocal scheme.

\begin{figure}[t]
	\vspace{-6mm}
	\centering
	\includegraphics[width=2.8in]{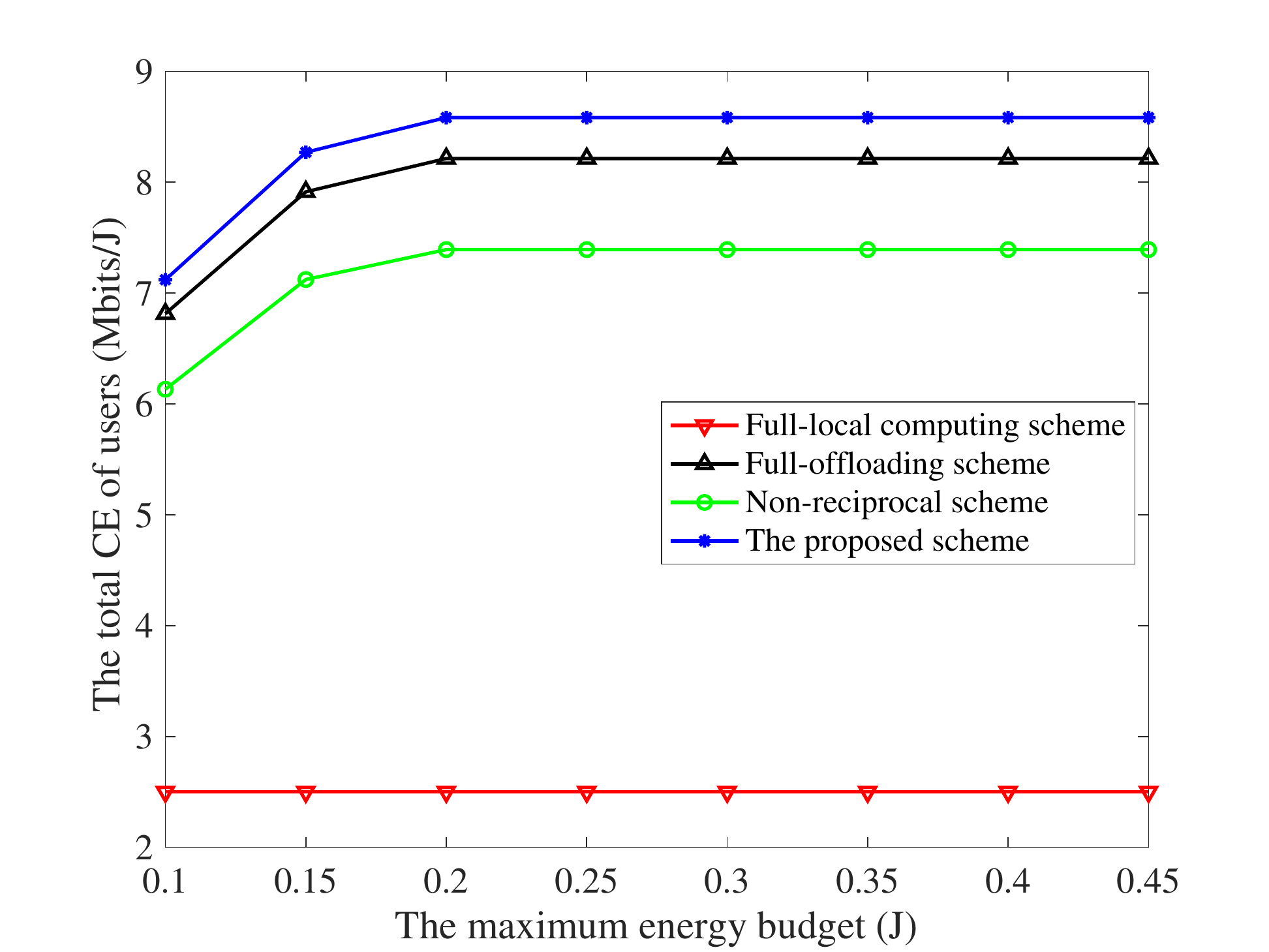}
	\caption{The total CE versus the maximum energy budget. }
	\label{fig2}
\end{figure}

Fig. \ref{fig2} shows the total CE versus the maximum energy budget of each user ($E_k^{\text{bud}}$) with $R_k^{\text{th}}=0.2$Mbits. As can be seen, except for the full-local computing scheme, the CE of the other schemes gradually increases with the increasing $E_k^{\text{bud}}$ and then stabilizes. This behavior is due to the fact that, as $E_k^{\text{bud}}$ gradually increases, not only the energy available for performing tasks ($E_k^{\text{tot}}$) increases, but also the number of computed bits ($R_k^{\text{tot}}$) increases. When the increasing trend of $R_k^{\text{tot}}$ exceeds that of $E_k^{\text{tot}}$, the total CE gradually increases until it reaches a stable balance point. This is because when using more energy to execute the task, the consumed energy gradually outweighs the enhancement brought by computed bits, resulting in a decrease in CE. This can also explain the variation tendency of the full-local computing scheme.

\begin{figure}[t]
	\vspace{-6mm}
	\centering
	\includegraphics[width=2.8in]{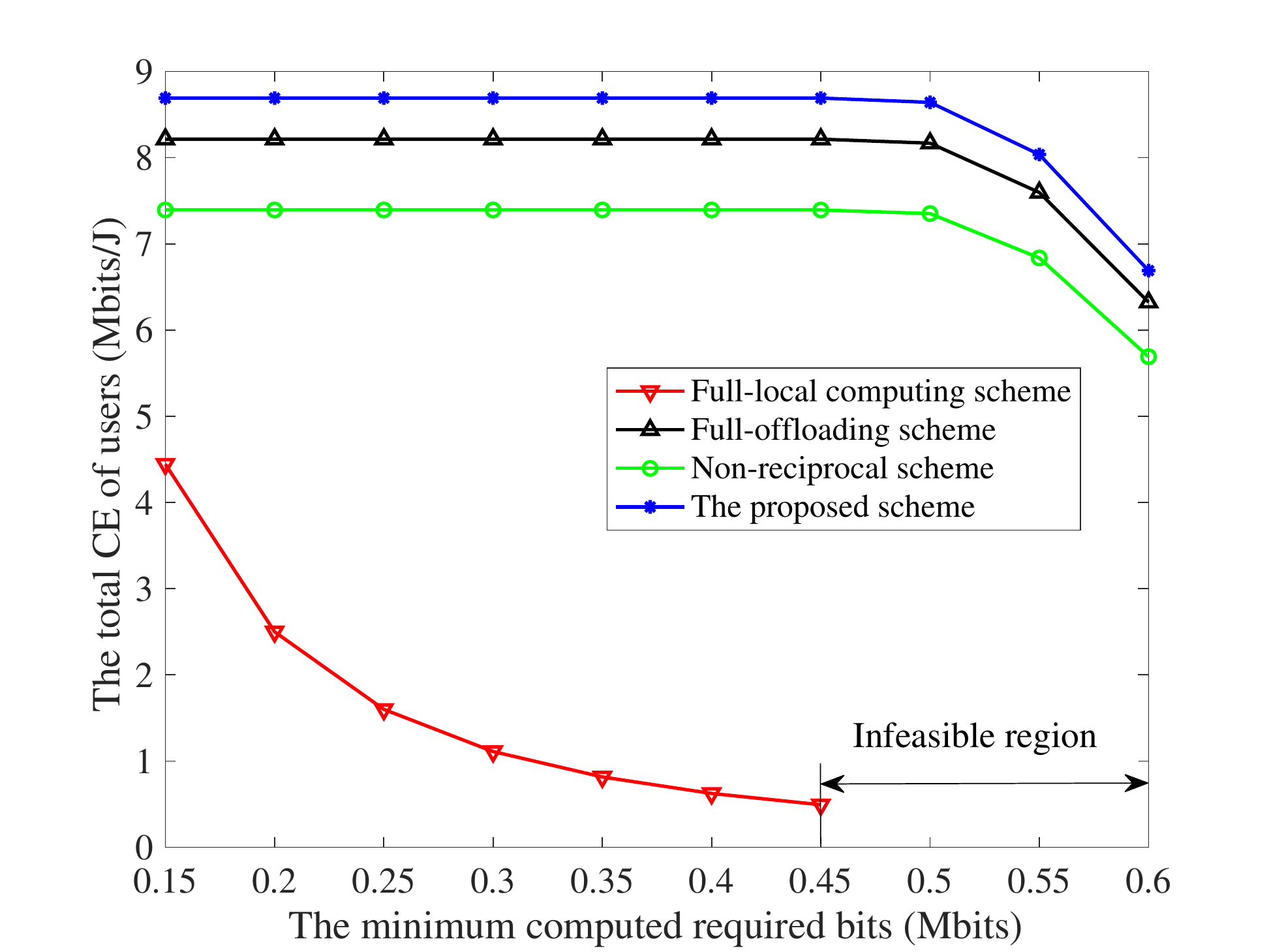}
	\caption{The total CE versus the minimum required computed bits. }
	\label{fig3}
\end{figure}

Fig. \ref{fig3} shows the total CE versus the minimum required computed bits of each user ($R_k^{\text{th}}$) with $E_k^{\text{bud}}=1$J. As shown, the total CE of all schemes exhibits an overall downward trend. Specifically, all schemes first remain unchanged and then decrease as $R_k^{\text{th}}$ increases, with the exception of the full-local computing scheme. This is because, as shown in (\ref{l1}) and (\ref{s7}), the total CE of the full-local computing scheme is inversely proportional to $R_k^{\text{l}}$, while $R_k^{\text{l}}$ cannot be less than $R_k^{\text{th}}$. Therefore, when $R_k^{\text{th}}$ increases, the total CE continues to decrease. Moreover, when $R_k^{\text{th}}$ is relatively large, more power is required to meet the quality of experience/service of users, resulting in significant energy consumption that exceeds the energy consumption corresponding to the steady CE, thereby reducing the total CE of other schemes. Moreover, the full-local computing scheme fails when $R_k^{\text{th}}$ is higher, as the resulting energy consumption exceeds the given energy budget.

Besides, from the above figures, the performance of the proposed scheme is better than that of other schemes, as partial offloading provides an efficient task allocation framework for task execution, and the addition of BackCom provides additional transmission capacity for users.

\section{Conclusions}

In this letter, we investigate and analyze resource allocation for the BackCom-assisted MEC system by taking the user reciprocity into account. The objective is to maximize the CE of users by jointly optimizing the transmit power, the offloading time, the RC, and the local computing frequency. An alternating algorithm is designed with  Dinkelbach's method and quadratic transform. Besides, the closed-form solutions of corresponding variables and the feasible condition for the reciprocal mode are further provided. Simulation results show that the proposed scheme is better than benchmark schemes in terms of the CE.

	% that's all folks
\end{document}